\documentclass[iop]{emulateapj}
\usepackage{hyperref}
\usepackage{color}
\usepackage{natbib}
\usepackage{graphicx}
\usepackage{xfrac}
\usepackage{ulem}
\usepackage{threeparttable}
\begin{document}
\title{The circumgalactic medium of \MakeLowercase{e}BOSS emission line galaxies: \\ signatures of galactic outflows in gas distribution and  kinematics}
\shorttitle{The CGM of ELGs}
\shortauthors{Lan \& Mo}
\author{Ting-Wen Lan\altaffilmark{1} and Houjun Mo\altaffilmark{2,3}}

\altaffiltext{1}{Kavli IPMU, the University of Tokyo (WPI), Kashiwa 277-8583, Japan}
\altaffiltext{2}{Physics Department and Center for Astrophysics, Tsinghua University, Beijing 10084, China}
\altaffiltext{3}{Department of Astronomy, University of Massachusetts, LGRT-B619E, 710 North Pleasant Street, Amherst, MA, 01003, USA}

\begin{abstract}
We study the distribution and kinematics of the cool circumgalactic medium (CGM) of 
emission line galaxies (ELGs) traced by metal absorption lines. Using about $200,000$ 
ELGs from SDSS-IV eBOSS and half a million background quasars from SDSS, we measure the 
median absorption strength of MgII and FeII lines in quasar spectra for impact 
parameters ranging from 10 kpc to 1 Mpc. For comparison we measure the same quantity 
around luminous red galaxies (LRGs). On scales greater than 100 kpc both ELGs and LRGs 
exhibit similar absorption profiles. However, metal absorption is 5-10 times stronger 
around ELGs on smaller scales. 
The metal absorption strength is anisotropic, with an 
excess along the minor-axis of the galaxies, indicating an outflow origin of the 
absorbing gas. 
The ratio between the velocity dispersion of the cool CGM 
and that of its host dark matter halo is about one for ELGs but about 
half for LRGs. These results show that the dichotomy of galaxy types 
is reflected in both the density distribution and kinematics of 
the CGM traced by metal absorption lines. Our results provide strong 
evidence that the CGM of ELGs is enriched by gas outflows generated by star 
formation. 
\end{abstract}

%
%
\section{Introduction}

Gas around galaxies, the circumgalactic media (CGM), contains 
information about gas accretion and outflows, important 
processes that drive the evolution of galaxies \citep[][for a review]{Tumlinson2017}. By probing the 
properties of the CGM and their connections with galaxies, one can 
hope to understand the influences of these mechanisms, thereby 
better understanding the formation and evolution of galaxies
in general. To this end, absorption line spectroscopy has been a 
powerful tool to extract the properties of the CGM through its absorption 
line signatures in the spectra of background objects. 

Since the first discovery of a pair of galaxy and metal absorber 
produced by the CGM \citep{Bergeron1986}, 
many investigations 
about the relationships between galaxies and their surrounding gas 
have been carried out over a wide range of redshift and using samples 
of at most a few hundred
galaxy-absorber pairs that are spectroscopically confirmed  
\citep[e.g.,][]{Steidel1994,Churchill2005,Chen2010,Steidel2010, 
Tumlinson2011,Nielsen2013,Bordoloi2014, Liang2014, Schroetter2016, Borthakur2016, Burchett2016,
Ho2017,Heckman2017,Johnson2017,Rubin2018,Lopez2018}.
On the other hand, large survey 
datasets offer the opportunity to explore the connection between 
galaxies and the CGM statistically  
\citep[e.g.,][]{Zibetti2007, Bordoloi2011, Menard2011, Zhu2013a, Lan2014, Zhu2014, Peek2015,Huang2016}. 
However, although interesting results have been obtained from these studies,
the properties of the gas around galaxies remain poorly constrained. 
In particular, a systematic investigation about the gas properties as a 
function of galaxy properties, which is required in order to understand 
the origin of the CGM, is still lacking. 

In this paper, we intend to characterize the properties of gas around 
star-forming galaxies at redshift $\sim 1$ by making use of the largest 
emission line galaxy (ELGs) catalog observed by the Extended Baryon Oscillation 
Spectroscopic Survey \citep[eBOSS,][]{Dawson2016} in Sloan Digital Sky Survey IV 
\citep[SDSS IV,][]{Blanton2017}. To this end, we cross-correlate the 
flux-decrement in the background quasar spectra with the presence of ELGs, 
and obtain the radial distribution and kinematics of the 
cool circumgalactic gas of star-forming galaxies. The structure of 
the paper is as follows. Our data analysis is described in Section 2, and 
our results are presented in Section 3. The implications of our results 
are discussed in Section 4, and we summarize in Section 5. 
Throughout the paper we adopt a flat $\Lambda$CDM cosmology with $h=0.7$ and 
$\Omega_{\rm M}=0.3$.

\begin{figure*}
\center
\includegraphics[width=1.02\textwidth]{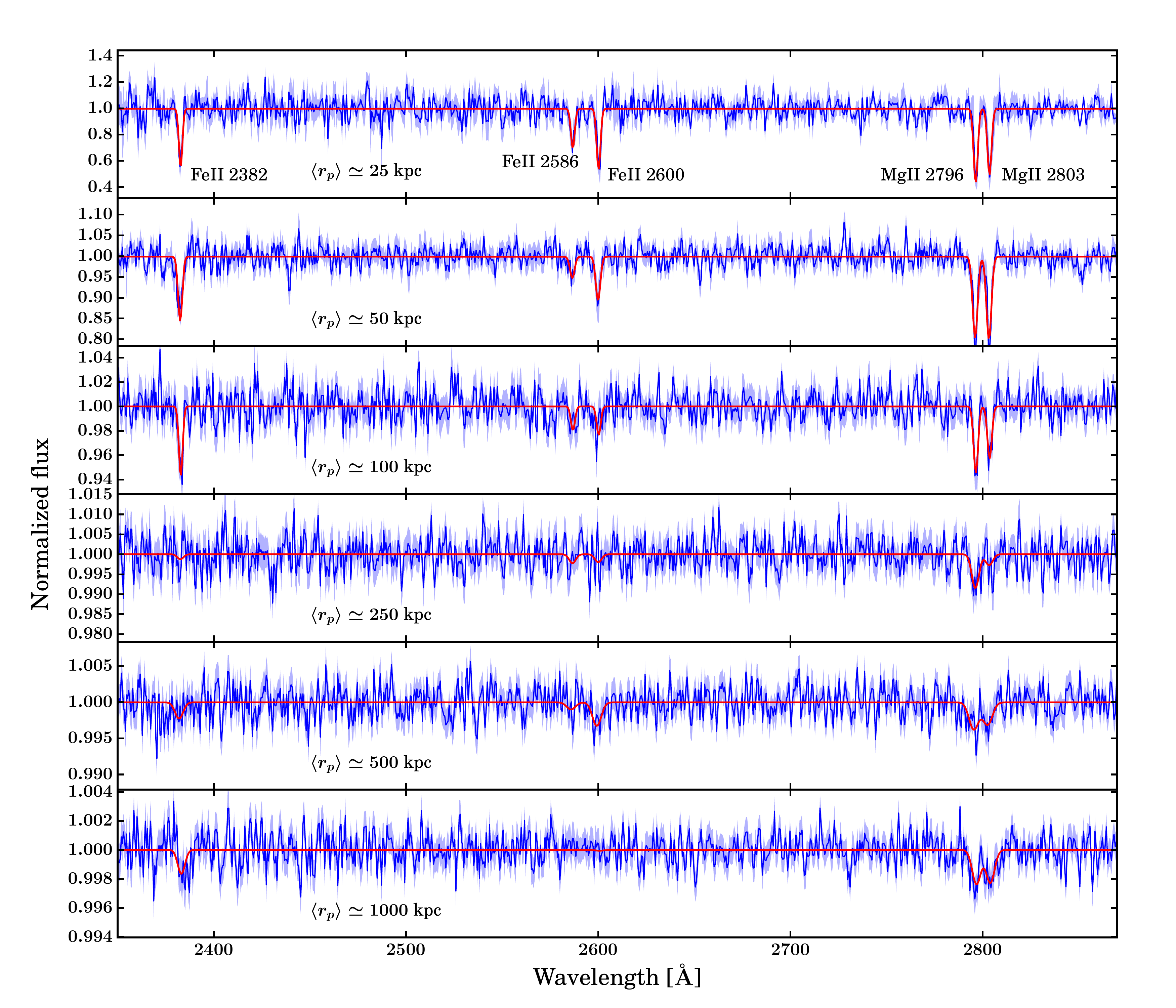}
\caption{Example of median composite spectra of background quasars at rest-frame of ELGs as a function of $r_{p}$. The spectra and the best-fit  MgII/FeII absorption lines are shown with blue and red solid lines, respectively. The blue shaded regions illustrate the bootstrapping uncertainties of the spectra. Note that the scale of y-axis changes as a function of $r_{p}$. The noise level of the composite spectrum at $\sim 1000$ kpc is about $10^{-3}$ of the continuum.}
\label{}
\end{figure*}
\begin{table*}[ht] 
\caption{MgII rest equivalent widths around ELGs (upper part) and LRGs (lower part)}
\centering
\begin{threeparttable}
  \begin{tabular}{ccccccccc}
\hline\hline 
$r_{p}$ bin [kpc] & $\langle r_{p}\rangle$ & $\rm N_{spec}$ & $\langle W_{\rm MgII} \rangle$ [\AA] &  $\sigma(\langle W^{\rm MgII} \rangle)$ & $\langle W_{\lambda2796} \rangle$ [\AA] &  $\sigma(\langle W_{\lambda2796} \rangle)$ & $\langle W_{\lambda2803} \rangle$ [\AA] &  $\sigma(\langle W_{\lambda2803} \rangle)$ \\ [1mm]

\hline\hline
$(10,28]$ & $23$ & $11$ & $2.830$ & $0.243$ & $1.529$ & $0.135$ &  $1.301$ & $0.142$ \\
$(28,40]$ & $34$ & $27$ & $2.491$ & $0.372$ & $1.349$ & $0.246$ &  $1.142$ & $0.182$ \\
$(40,56]$ & $50$ & $50$ & $1.307$ & $0.175$ & $0.647$ & $0.121$ &  $0.660$ & $0.080$ \\
$(56,78]$ & $69$ & $105$ & $0.622$ & $0.106$ & $0.355$ & $0.066$ &  $0.267$ & $0.065$ \\
$(78,110]$ & $95$ & $183$ & $0.287$ & $0.050$ & $0.161$ & $0.033$ &  $0.127$ & $0.029$ \\
$(110,155]$ & $134$ & $406$ & $0.113$ & $0.045$ & $0.056$ & $0.024$ &  $0.057$ & $0.034$ \\
$(155,218]$ & $190$ & $774$ & $0.030$ & $0.030$ & $0.023$ & $0.021$ &  $0.006$ & $0.027$ \\
$(218,307]$ & $266$ & $1571$ & $0.052$ & $0.023$ & $0.040$ & $0.014$ &  $0.012$ & $0.016$ \\
$(307,431]$ & $375$ & $2946$ & $0.044$ & $0.019$ & $0.017$ & $0.015$ &  $0.027$ & $0.016$ \\
$(431,606]$ & $527$ & $5902$ & $0.044$ & $0.013$ & $0.024$ & $0.010$ &  $0.019$ & $0.009$ \\
$(606,853]$ & $739$ & $11512$ & $0.022$ & $0.010$ & $0.016$ & $0.007$ &  $0.006$ & $0.007$ \\
$(853,1200]$ & $1040$ & $22999$ & $0.025$ & $0.006$ & $0.013$ & $0.004$ &  $0.012$ & $0.004$ \\
\hline
$(10,28]$ & $22$ & $46$ & $0.500$ & $0.155$ & $0.273$ & $0.098$ &  $0.227$ & $0.098$ \\
$(28,40]$ & $35$ & $75$ & $0.447$ & $0.193$ & $0.234$ & $0.127$ &  $0.214$ & $0.112$ \\
$(40,56]$ & $48$ & $157$ & $0.550$ & $0.093$ & $0.278$ & $0.062$ &  $0.272$ & $0.050$ \\
$(56,78]$ & $68$ & $362$ & $0.250$ & $0.052$ & $0.150$ & $0.031$ &  $0.100$ & $0.028$ \\
$(78,110]$ & $95$ & $670$ & $0.290$ & $0.043$ & $0.181$ & $0.031$ &  $0.108$ & $0.027$ \\
$(110,155]$ & $135$ & $1396$ & $0.135$ & $0.027$ & $0.078$ & $0.015$ &  $0.057$ & $0.017$ \\
$(155,218]$ & $189$ & $2665$ & $0.100$ & $0.023$ & $0.060$ & $0.017$ &  $0.040$ & $0.015$ \\
$(218,307]$ & $266$ & $5542$ & $0.098$ & $0.014$ & $0.053$ & $0.009$ &  $0.045$ & $0.008$ \\
$(307,431]$ & $380$ & $12253$ & $0.053$ & $0.011$ & $0.031$ & $0.009$ &  $0.022$ & $0.010$ \\
$(431,606]$ & $526$ & $28672$ & $0.041$ & $0.010$ & $0.024$ & $0.006$ &  $0.017$ & $0.005$ \\
$(606,853]$ & $739$ & $56189$ & $0.028$ & $0.006$ & $0.015$ & $0.004$ &  $0.012$ & $0.004$ \\
$(853,1200]$ & $1041$ & $110921$ & $0.018$ & $0.004$ & $0.009$ & $0.002$ &  $0.009$ & $0.002$ \\
\hline
\end{tabular}
\end{threeparttable}
\label{table:all_data}
\end{table*}
\begin{figure*}
\center
\includegraphics[width=0.9\textwidth]{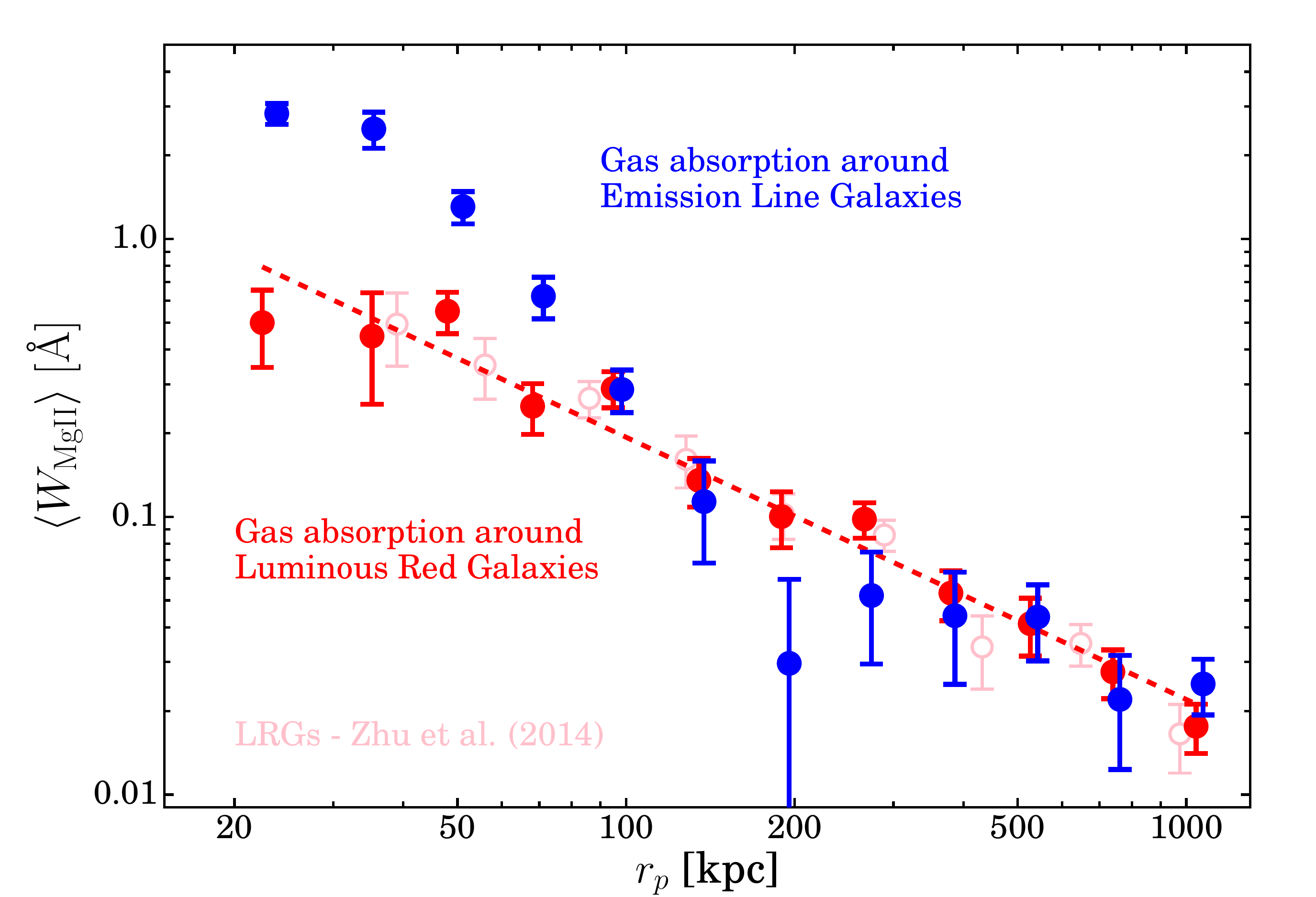}
\caption{MgII rest equivalent width, the sum of MgII$\lambda 2796$ and $\lambda 2803$ lines, as a function of $r_{p}$. The gas 
profiles of ELGs and LRGs are shown with blue and red data points, respectively. 
The pink open circles show the LRG measurements from \citet{Zhu2014}.
}
\label{}
\end{figure*}

%
%
\section{Data analysis}
We study the properties of the CGM traced by metal absorption lines imprinted in the
background quasar spectra. Our analysis is based on two spectroscopic galaxy samples 
provided by SDSS as foreground galaxies: (1) emission line galaxies (ELGs) 
and (2) luminous red galaxies (LRGs), using all the SDSS quasars as background 
objects.  In the following, we describe in detail the samples and method for our analysis. 
\subsection{Datasets}

\textbf{Emission line galaxies -} We use the emission line galaxy (ELGs) catalog compiled 
from the SDSS-IV eBOSS survey \citep{Blanton2017, Dawson2016}. This survey aims to 
detect star-forming galaxies with strong emission lines in the redshift range 0.8-1.0 and use them
as tracers of the large-scale structure to investigate the baryonic acoustic 
oscillations. The galaxies are targeted from images obtained by the DECam Legacy 
Survey \citep{Dey2018} \footnote{\url{http://legacysurvey.org/}} with photometric selections 
that maximize the survey efficiency \citep{Comparat2016,Raichoor2016}. The 
spectra of the ELGs are obtained by the BOSS spectrograph \citep{Smee2013}  on the APO 2.5-meter SDSS telescope \citep{Gunn2006} and processed by the SDSS spectroscopic pipeline 
\citep{Bolton2012}\footnote{Version 5-10-7}, which automatically estimates the redshifts 
of galaxies and derives the properties of emission lines \citep{Hutchinson2016}.  

We select ELGs with redshifts greater than 0.4 with reliable redshift estimation \citep{Comparat2016}. 
This yields a total of about 180,000 ELGs 
in the redshift range between $0.4$ and $1.5$, with a median value 
about $0.85$. The typical uncertainty of the redshift estimate is about 
20 km/s.

The ELGs have a typical stellar mass of about $3 \times 10^{10} \, M_{\odot}$ 
\citep{Raichoor2017} and are hosted by dark matter halos with a typical mass 
of about $1.5\times10^{12}\, M_{\odot}$ \citep{Favole2016}. 
Based on their [OII] $\lambda 3727$ luminosities and the conversion model 
given by \citet{Kennicutt1998}, the ELGs have star formation rates 
(SFR) ranging from $1$ to $20 \, {\rm M_{\odot}/yr}$, with a median value 
of about $8 \, {\rm M_{\odot}/yr}$. We note that the [OII] luminosities are not corrected for dust reddening.
 
\textbf{Luminous red galaxies -}
We use the sample of luminous red galaxies (LRGs) provided by SDSS DR14 
data release \citep{Abolfathi2017} from the BOSS \citep{Dawson2013} 
and eBOSS \citep{Prakash2016, Dawson2016} surveys. 
We select LRGs with redshifts greater than 0.4 
and with redshift uncertainties smaller than 35 km/s. This yields a sample 
of about 760,000 LRGs, with median redshift of about 0.55. The typical 
stellar and halo masses of the LRGs are $\sim 10^{11.2} \rm \, M_{\odot}$ 
and $\sim 10^{13.5} \rm \, M_{\odot}$, respectively 
\citep[e.g.,][]{White2011,Zhai2017}. We note that while the MgII absorption 
around LRGs has been measured by \citet{Zhu2014} with a method
similar to the one used here, we repeat the analysis with the latest 
sample (1) to obtain the FeII absorption which is not provided by \citet{Zhu2014}, and (2) as a consistent check between the new data
and previous ones. 

\textbf{Background quasars -} 
For background objects, we use the SDSS DR14 quasar catalog 
\citep{Abolfathi2017,Paris2017}, which contains all the quasars observed by 
the SDSS surveys. We select quasar-galaxy pairs that have projected distances 
smaller than $1.2 \, \rm Mpc$ and have the quasar redshift higher 
than galaxy redshift by at least $0.1$. This selection 
yields in total about 70,000 ELG-quasar pairs and 320,000 LRG-quasar pairs.

\subsection{Method}
We measure the average MgII and FeII absorption strengths 
(tracers of cool gas with $T\sim10^{4}$ K)
in the quasar spectra as
functions of the impact parameter, $r_{p}$. defined to be the distance 
between the galaxy and the line of sight to its paired quasar.  
We follow the procedure in \citet{Zhu2013} to estimate the absorption line 
strengths. We first model and remove spectral features in the quasar 
spectra that are intrinsic to quasars, using the quasar eigen-spectra 
provided by \citet{Zhu2013}, together with a dimensional reduction technique
, called non-negative matrix factorization \citep[NMF,][]{Lee2001}, 
as implemented by \citet{Zhu2016}\footnote{The code can be found at 
\url{https://github.com/guangtunbenzhu/NonnegMFPy}.}. 
Intermediate-scale fluctuations are removed by a median filter with 
a width of $71$ pixels. We also remove the systematic features originated 
from the SDSS pipeline in the observer frame \citep{Lan2018}.
Finally, we combine the continuum-normalized spectra of background quasars 
at the rest-frame of the foreground galaxies to obtain the composite spectra,
which have higher S/N than the original individual spectra. 
A median estimator is used to avoid the impacts of outliers. 
We found that using a robust mean estimator yields consistent results. 
To make the composite spectra, we use pixels of quasar spectra with S/N 
greater than 3; including pixels with lower S/N does not enhance the 
quality of the final composite spectra. With this S/N selection, the composite spectra are made with $\sim70\%$ of total galaxy-quasar pairs effectively. 
A similar method has been used in 
the analyses of gas absorption at both high \citep{Steidel2010} and 
low \citep{Zhu2013a} redshifts, as well as using galaxies as background 
sources \citep{Bordoloi2011}.

Figure 1 shows examples of the median composite spectra as a function 
of the impact parameter from ELGs. The shaded regions show the uncertainties of 
the composite spectra from 200 bootstrap samples. Note that the 
scale of vertical axis changes with the impact parameter. 
Note also that the signal-to-noise ratio of the composite spectra 
per spectral element is about 1000 at $r_{p}\sim 1000\,{\rm kpc}$. 

We measure the rest equivalent widths of the MgII$\lambda\lambda$2796, 2803 and FeII$\lambda\lambda$2600, 2586, 2382 lines by fitting the spectra 
to Gaussian profiles with their amplitudes as free parameters. 
Since MgII and FeII lines trace gas clouds with the same physical properties and velocities \citep[e.g.,][]{Churchill2001,Churchill2003},
we fit all the line widths with a single velocity dispersion parameter.
The uncertainties of the 
rest-frame equivalent widths and velocity dispersion are estimated by 
bootstrapping the sample 200 times. Examples of the best-fit Gaussian 
profiles are shown with red solid lines in Figure 1.

MgII absorption lines have been detected individually in random quasar sightlines, 
and their rest equivalent width distribution is found to
follow roughly an exponential distribution \citep[e.g.,][]{Nestor2005,Zhu2013}. 
For a MgII absorber with rest equivalent width greater than 0.4 $\rm \AA$, 
the absorption line is mostly saturated and therefore, the rest equivalent width 
reflects a combination of the internal velocity dispersion of the system and its column density. The probability to detect these absorbers around galaxies, 
namely the covering fraction, depends on the galaxy population as well as on 
the impact parameter from the galaxies \citep[e.g.,][]{Chen2010, Nielsen2013,Lan2014}. 
Since our composite spectra are made by combining random background quasar spectra 
{\it without prior knowledge} about individual metal absorbers, the measured 
average rest equivalent width, $\langle W_{\rm MgII} \rangle$, reflects the 
summation of the product between the covering fraction and the absorption 
strength over individual MgII absorbers: 
\begin{equation}
    \langle W_{\rm MgII} \rangle = \sum_{i} 
    f_{c}(W^{\rm l}_{i}<W_{{\rm MgII},i}<W^{\rm u}_{i}, r_p) 
    \times\hat{W}_{{\rm MgII}, i},
\end{equation}
where $f_{c}(W^{\rm l}_{i}<W_{{\rm MgII},i}<W^{\rm u}_{i},r_p)$ is the covering 
fraction of absorbers with strength in $[W^{\rm l}_{i},W^{\rm u}_{i}]$
at a given impact parameter $r_p$, and $\hat{W}_{{\rm MgII},i}$ is the 
average strength. Thus, even if $\langle W_{\rm MgII} \rangle$ 
measured from our composite spectra appears to be unsaturated 
($W<0.4 \rm \, \AA$), it may be due to the combination of a 
low covering fraction with strongly saturated absorbers. This is consistent with the fact that the line ratio between MgII $\lambda2796$ and $\lambda2803$ 
is larger than $0.5$ at scales $>200$ kpc (see Fig. 1), even though 
the rest equivalent widths are small, suggesting that the average absorption 
is indeed dominated by strong saturated absorbers.
Consequently, a direct conversion from the measured rest equivalent width to 
a column density may result in an underestimation of the metal abundance. 
In the following, we will first present our results in terms of the median rest 
equivalent width obtained from the composite spectra (Section 3), and then 
discuss the covering fraction and gas mass inferred from them (Section 4).

%
%
\section{Results}

\subsection{Gas distribution traced by MgII/FeII absorption lines}

We measure the radial distribution of gas traced by MgII/FeII absorption 
lines. Figure 2 shows the median rest equivalent widths of the sum of 
MgII$\lambda$2796 and MgII$\lambda$2803 lines around ELGs and LRGs as functions of the impact parameter. The individual measurements are listed in Table 1 for reference.

It can be seen that the MgII absorption strength around both ELGs and LRGs 
decreases monotonically with $r_{p}$. For ELGs, there seems to be a change 
in slope at $r_{p}\sim 100 {\rm \, kpc}$: the absorption profile is roughly 
a power law, $W\propto r_{p}^{-1}$ at larger $r_p$,   
and the profile is steeper, $r_{p}^{-1.5}$, at smaller $r_{p}$. 
In contrast, the MgII absorption profile around LRGs, shown by 
the red data points, can be described by a single power-law, 
$W\propto r_{p}^{-\alpha}$ with $\alpha= -0.94\pm 0.05$, as illustrated 
by the red dashed line. 
The absorption profile around LRGs obtained here 
is consistent with the measurements of \citet{Zhu2014} (shown by the 
pink open circles), demonstrating the robustness of the observational results. In addition, our measurements extend the profile 
to $r_p<30$ kpc.

At $r_p>100 \, {\rm kpc}$ the gas absorption profile around ELGs is 
consistent with that around LRGs, but the amplitude is slight 
lower, by about $85\%$, albeit the large uncertainties. This difference 
in amplitude at $r_p>100\,{\rm kpc}$ may be due to that ELGs and LRGs 
reside in dark matter halos with average masses of $\sim10^{12} \, M_{\odot}$ and 
$\sim10^{13.5} \, M_{\odot}$, respectively. However, measurements with higher S/N are required to test this picture. 
At $r_p<100 \, {\rm kpc}$, the absorption profile around ELGs deviates significantly from the 
power law extrapolated from larger scales, and the absorption equivalent width
is about a factor of 5-10 as large as that around LRGs. Such a trend is also
observed in the FeII absorption profiles. 

If we scale $r_p$ by the corresponding halo virial radius, $R_v$, the 
small scale difference between ELGs and LRGs will be enhanced. 
This result indicates that the excess of MgII absorption around 
ELGs is limited in their host halos, likely due to gas outflows produced 
by the star formation activities in central galaxies
\citep[e.g.][]{Bordoloi2011,Lan2014}. 
This result demonstrates that the properties of galaxies is reflected in their surrounding cool gas. 

\begin{figure}
\includegraphics[width=0.48\textwidth]{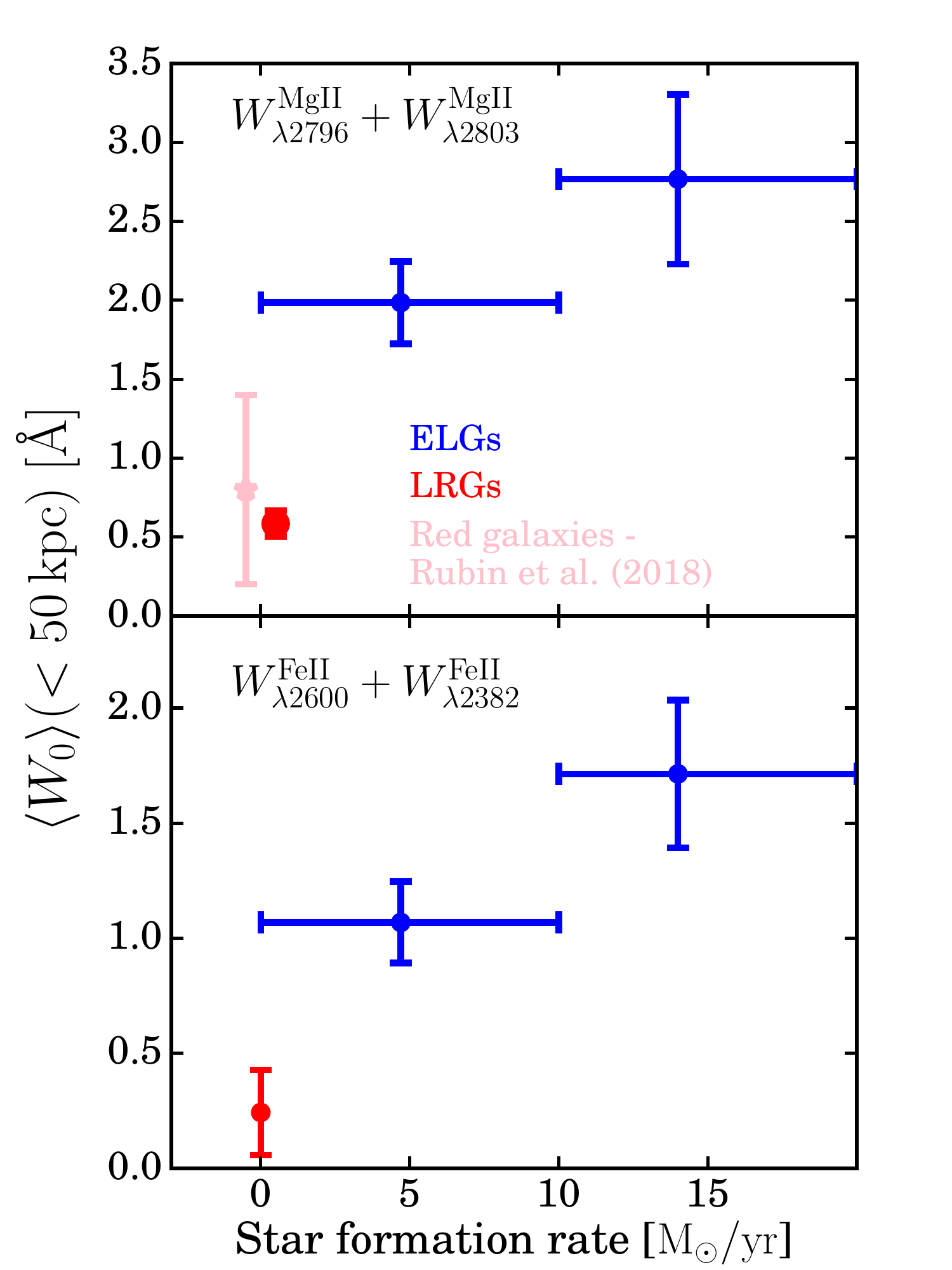}
\centering
\caption{Absorption properties as a function of star-formation rate 
obtained from systems with impact parameters within 50 kpc. 
{\it Top:} the total width of the $\rm MgII\lambda \, 2796$ and 
$\rm MgII\lambda \, 2803$ lines. 
The same measurements around LRGs and red galaxies \citet{Rubin2018} are shown in red and pink, respectively. 
{\it Bottom:} the total width of the $\rm FeII\lambda \, 2600$ 
and $\rm FeII\lambda \, 2382$ lines. The results show that both the MgII and
FeII absorption strengths are correlated with the SFR of the host galaxies.
}
\label{}
\end{figure}
\begin{figure*}
\centering
\includegraphics[width=0.9\textwidth]{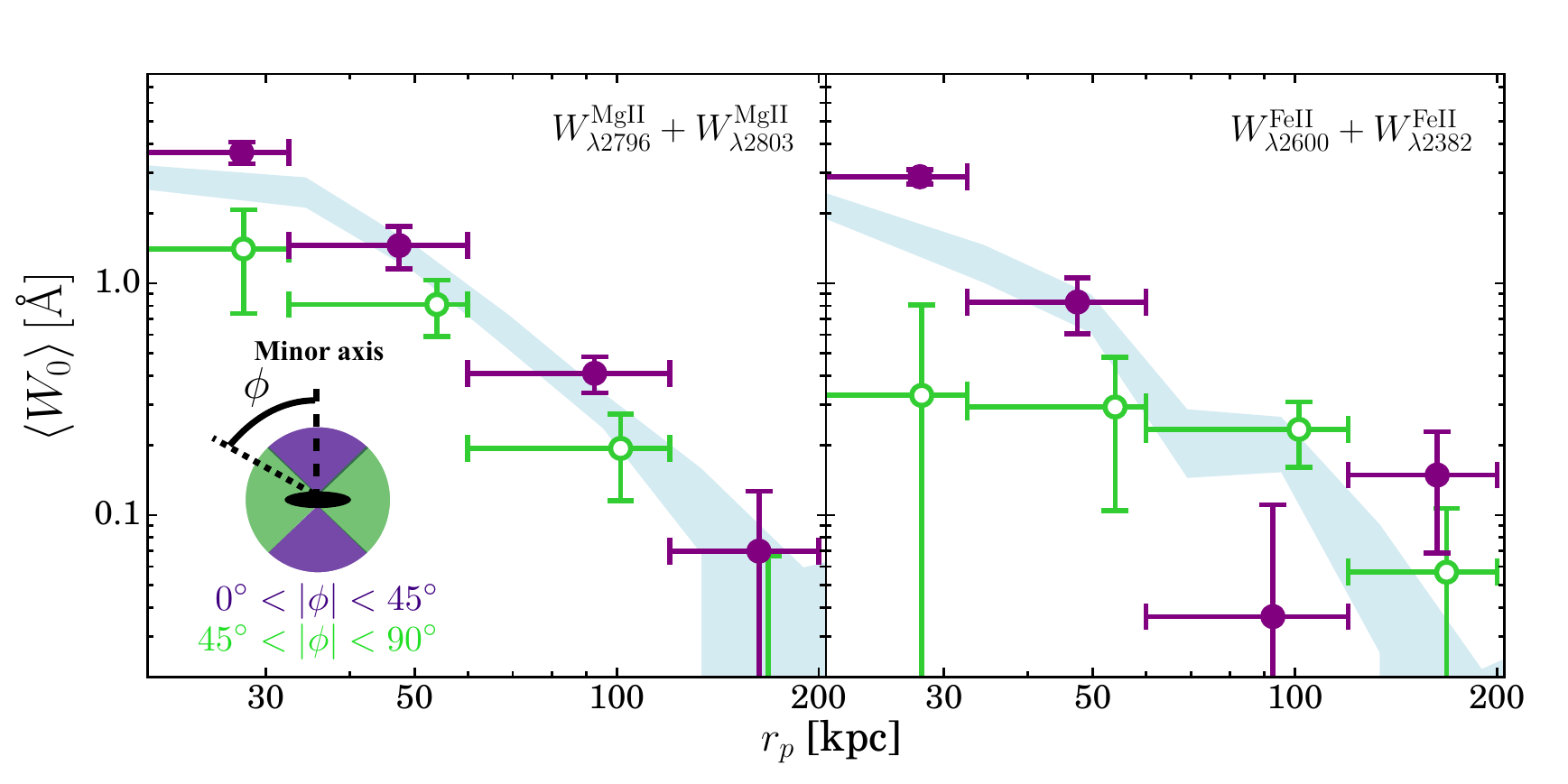}
\caption{Dependence of the absorption profile on the azimuthal angle, $\phi$, 
between the impact-parameter vector and the semi-minor axis of ELGs with exponential disk profiles. 
{\it Left:} the total width of the $\rm MgII\lambda \, 2796$ and 
$\rm MgII\lambda \, 2803$ lines. 
{\it Right:} the total width of the $\rm FeII\lambda \, 2600$ and 
$\rm FeII\lambda \, 2382$ lines.
The purple and green data points show the absorption strengths along minor and major axis respectively. The blue bands show the measurements obtained by 
averaging over all directions.
}
\label{fig:azimuthal}
\end{figure*}

\subsection{Correlation with star-formation rate}

To investigate how the absorption strength depends on the properties of 
galaxies, we separate the ELGs into two 
subsamples in SFR, one with $SFR >10 M_{\odot}/yr$ and the other 
$SFR < 10 M_{\odot}/yr$, and measure their absorption properties.
As shown in the two panels of Figure 3, the MgII and FeII absorption 
strengths at $r_{p}<50$ kpc both increase with SFR. 
Comparing the average absorption strengths around ELGs with 
that around LRGs and passive galaxies of similar stellar 
masses \citep{Rubin2018}, one can see that the trend with SFR 
extends to red galaxies that have very low star formation rates. 
Nevertheless, the absorption strengths of the two subsamples are 
consistent with each other at $r_{p}>50$ kpc.
This trend is consistent with previous results obtained by 
\citet{Bordoloi2011}, \citet{Lan2014}, and \citet{Rubin2018}.
Our result demonstrates that the correlation between the 
SFR of galaxies and the properties of the circumgalactic gas is 
reflected not only in the MgII absorption, but also in the FeII 
absorption in the inner region of the CGM.

\subsection{Dependence on azimuthal angle}

The dependence of the metal absorption profile on the azimuthal angle, $\phi$, 
defined to be the angle between the minor axis of the galaxy and the 
impact parameter vector of  the background quasar, provides further insights 
into the origin of the circumgalactic gas. For example, gas clouds 
driven by gas outflows from a star-forming galaxy 
are expected to be preferentially along the minor axis of the galaxy
\citep[][for a review]{Veilleux2005}, while cooling gas from a hot gaseous halo should have 
a more isotropic distribution. To investigate such dependence, we use the 
photometric information from DECaLS DR5 catalog \citep{Dey2018} and 
select ELGs that are best-fitted by exponential disk profiles as 
determined by {\it The Tractor}\footnote{\url{https://github.com/dstndstn/tractor}} \citep{Lang2016} \citep[See also Section 8 in][]{Dey2018}. 
This selection reduces the sample size by $\sim60\%$.
In short, the Tractor adopts a forwarding modeling approach by convolving modeled profiles with the point spread function of each individual exposure and finds the best-fit model profile that minimizes the residuals of all images. 
The algorithm only classifies extended objects into exponential disk profiles 
when the exponential disk profiles improve the fits by more than 
3$\sigma$ in comparison to round exponential profiles.
We use the image shape parameters provided by {\it the Tractor} 
to estimate the azimuthal 
angles for individual galaxy-quasar pairs and the corresponding 
uncertainties, which are typically 5-10 degrees.

Figure~\ref{fig:azimuthal} shows the MgII and FeII absorption strength as a 
function of $r_{p}$ in two azimuthal angle bins, one with 
$\phi\in [0^\circ, 45^\circ)$ and the other with 
$\phi\in [45^\circ, 90^\circ]$ (see the cartoon plotted 
in the left panel for the relevant geometry).  
The SFR distributions of ELGs in the two bins are consistent with each other. 
As one can see, the MgII and FeII absorption strengths at $r_{p}<50$ kpc
are about 2 times larger along the minor axis of ELGs (purple) than along 
the major axis (green). At larger $r_p$, the absorption strengths 
for the two azimuthal angle bins are comparable. 
For reference, the blue bands show the measurements by 
averaging over all directions. As expected, the averages go through 
the measurements of the two azimuthal bins. 
We also perform similar 
measurements around LRGs and find no dependence on the azimuthal angle, 
consistent with the finding of \citet{Huang2016}. 

This orientation dependence of the MgII/FeII absorption around ELGs demonstrates 
that a significant fraction of the cool gas traced by the absorbers  
is likely contributed by outflows originated from the galaxies (disks). This result is 
consistent with those obtained earlier using smaller data sets from the HST, 
COSMOS, Keck, VLT and SDSS \citep[e.g.,][]
{Bordoloi2011, Bouche2012, Kacprzak2012, Lan2014}. 
Other mechanisms, such as gas inflows and/or gas associated with satellite 
galaxies, are not expected to produce the azimuthal dependence observed, as we will discuss in \S\ref{ssec_origins}.

Our results show that the metal absorption strength around ELGs is anisotropic. Based on 
the fact that the absorption strength along the minor axis is about two times 
stronger than that along the major axis, we infer that about $2/3$ of 
the absorption is along the minor axis. Under the assumption that the metal 
absorption observed with $\phi\in [0^\circ, 45^\circ)$ is all due to outflows,  
this suggests that outflow gas contributes about 2/3 of the gas around 
ELGs within 50 kpc. This value should be considered as a lower limit, given 
that the opening angles of outflows may be larger than $45^{\circ}$. Moreover, the 
intrinsic azimuthal angle dependence is expected to be stronger than our 
measurements indicate,  because galaxy shapes estimated from the ground-based 
observation may be affected by seeing. In the near future, with the imaging 
data from the Euclid survey \citep{Amiaux2012}, one will be able to 
measure the azimuthal dependence of the gas absorption in more detail.

\subsection{Gas Kinematics}
\label{ssec_kinematics}

In the composite spectra, the metal absorption lines in concern 
are contributed by multiple clouds 
with a range of velocities relative 
to the host galaxies. The line widths, therefore, reflect the line-of-sight 
velocity dispersion of the gas, $\sigma_{\rm gas}$, around the galaxies, which is a convolution between the internal velocity dispersion of absorption systems and the relative velocity difference between the absorbers and the galaxies.
In Figure 5, we show the 
best-fit velocity dispersion 
(obtained by a joint fit of 5 metal lines) of MgII/FeII gas clouds around ELGs as a function of 
the impact parameter, with effects produced by the SDSS spectral resolution
and by the uncertainty in galaxy redshift
subtracted in quadrature. The measured
velocity dispersion of gas 
around ELGs is about $100 \,{\rm km/s}$ from $r_p=20$ to 100 kpc, and the 
results at larger $r_p$ become uncertain. 
The blue shaded region shows the $1\sigma$ range of the best-fit gas 
velocity dispersion obtained by assuming a constant velocity dispersion.
For comparison, the velocity dispersion of the gas around LRGs is shown 
in red.  
We note that our measured gas velocity dispersion around LRGs is 
consistent with those of \citet{Zhu2014} and \citet{Huang2016}. 
The gas velocity dispersion around LRGs is about 60 km/s larger than that  
around ELGs. 

\begin{figure}
\includegraphics[width=0.47\textwidth]{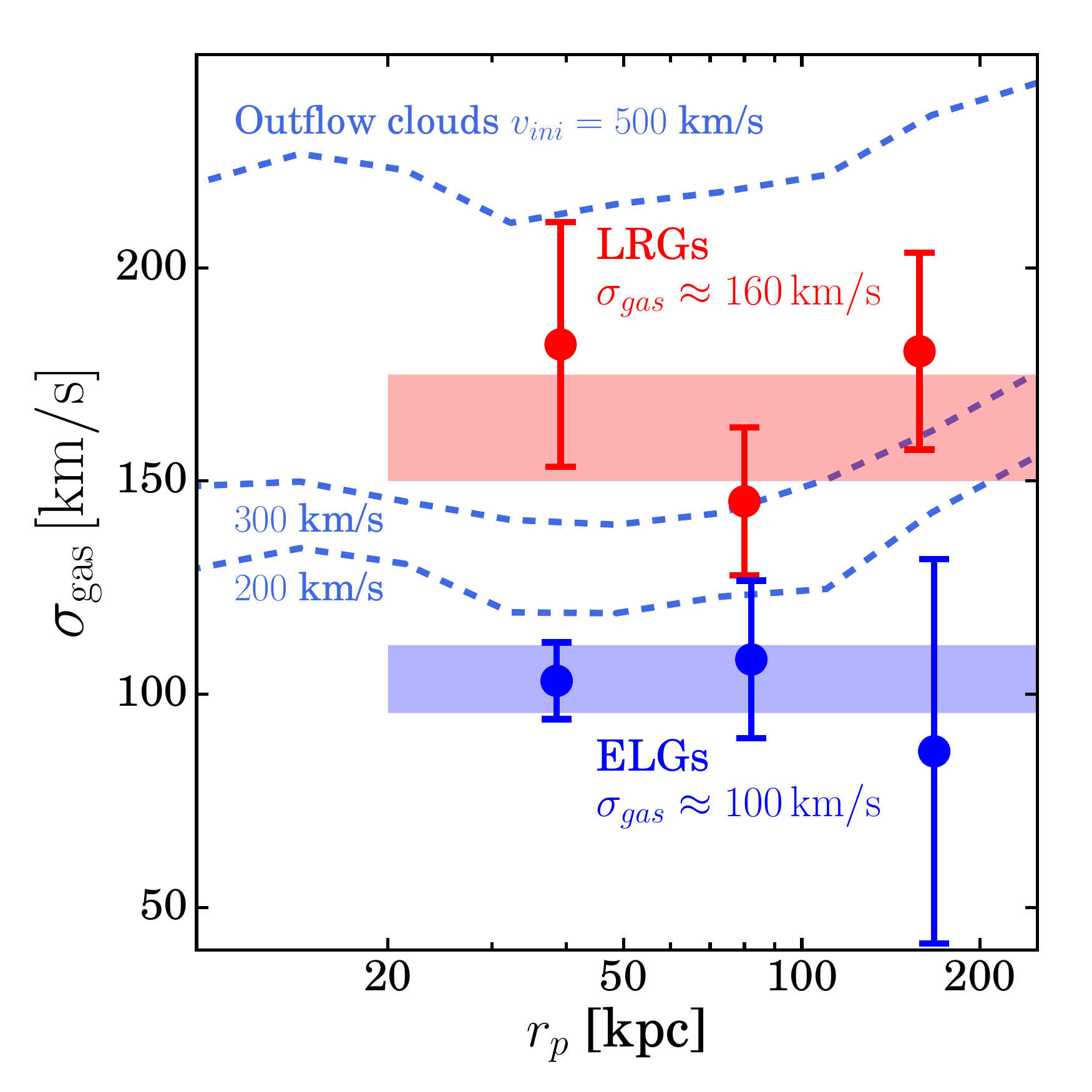}
\caption{Line-of-sight gas velocity dispersion, $\rm \sigma_{gas}$, as a function of $r_{p}$.
The $\sigma_{gas}$ around ELGs is about 100 km/s from 20 to 200 kpc, as shown by the blue data 
points. 
The blue shaded region shows the $1 \sigma$ range of the best-fit $\sigma_{gas}$. The same quantity for LRGs is shown in red. 
The blue dashed lines show the expected velocity dispersion of gas particles based on a simple
spherical outflow model described in the main text.}
\label{}
\end{figure}

Since ELGs reside in smaller dark matter halos than LRGs, it is informative 
to compare the gas velocity dispersion with that expected from the halo    
gravitational potential wells. Figure 6 shows the velocity dispersion of gas 
around galaxies as a function of halo mass. In addition to the measurements for ELGs and LRGs, shown by the blue and red data points, respectively, we also plot the dark matter velocity dispersion as a function of halo mass \citep[e.g.,][]{Elahi2018}, 
\begin{equation}
    \sigma_{m} \simeq 430 \, \bigg(\frac{M_{halo}}{10^{14} \, M_{\odot}}\bigg)^{1/3} \rm km/s,
\end{equation}
as the blue dashed line. The velocity dispersion of the gas around 
ELGs is consistent with that of dark matter in their halos, while 
the gas velocity dispersion around LRGs is only about half of the $\sigma_{m}$, 
as shown by the red dotted line. Thus, we have gas velocity bias
\begin{equation}
{\sigma_{gas}\over \sigma_{m}}\sim 
\left\{\begin{array}{lr}
 1    & \text{for ELGs;}\\
 0.5  & \text{for LRGs.}
 \end{array}
\right.
\end{equation}
A similar correlation between the kinematics of gas and galaxy types is also found in \citet{Nielsen2015} 
and \citet{Nielsen2016}, where the authors measure the velocity dispersion of gas clouds relative to the column-density weighted median velocity. Figure 6 shows the velocity dispersion of MgII gas clouds around blue and red galaxies, derived from about 30 galaxy-absorber pairs given by \citet{Nielsen2015} and \citet{Nielsen2016}, 
with the uncertainty estimated from bootstrapping. The halo masses are estimated 
from an abundance matching method \citep[e.g.,][]{Conroy2006}. Together with their data, the trend of gas velocity bias extends to even smaller halos. However, 
the methods and datasets used here and in \citet{Nielsen2015,Nielsen2016}
are quite different, and so the comparisons may be uncertain. 
Clearly, a homogeneous dataset covering a wide range of halo masses and 
galaxy types is required to quantify the gas velocity bias as a function 
of halo mass. Nevertheless, our result shows clearly that the 
$\sigma_{gas}$ of gas around star-forming and passive galaxies 
have very different behaviors with respect to that of dark matter, 
indicating that multiple mechanisms are  
driving the motion of the circum-galactic gas.

\begin{figure}
\includegraphics[width=0.47\textwidth]{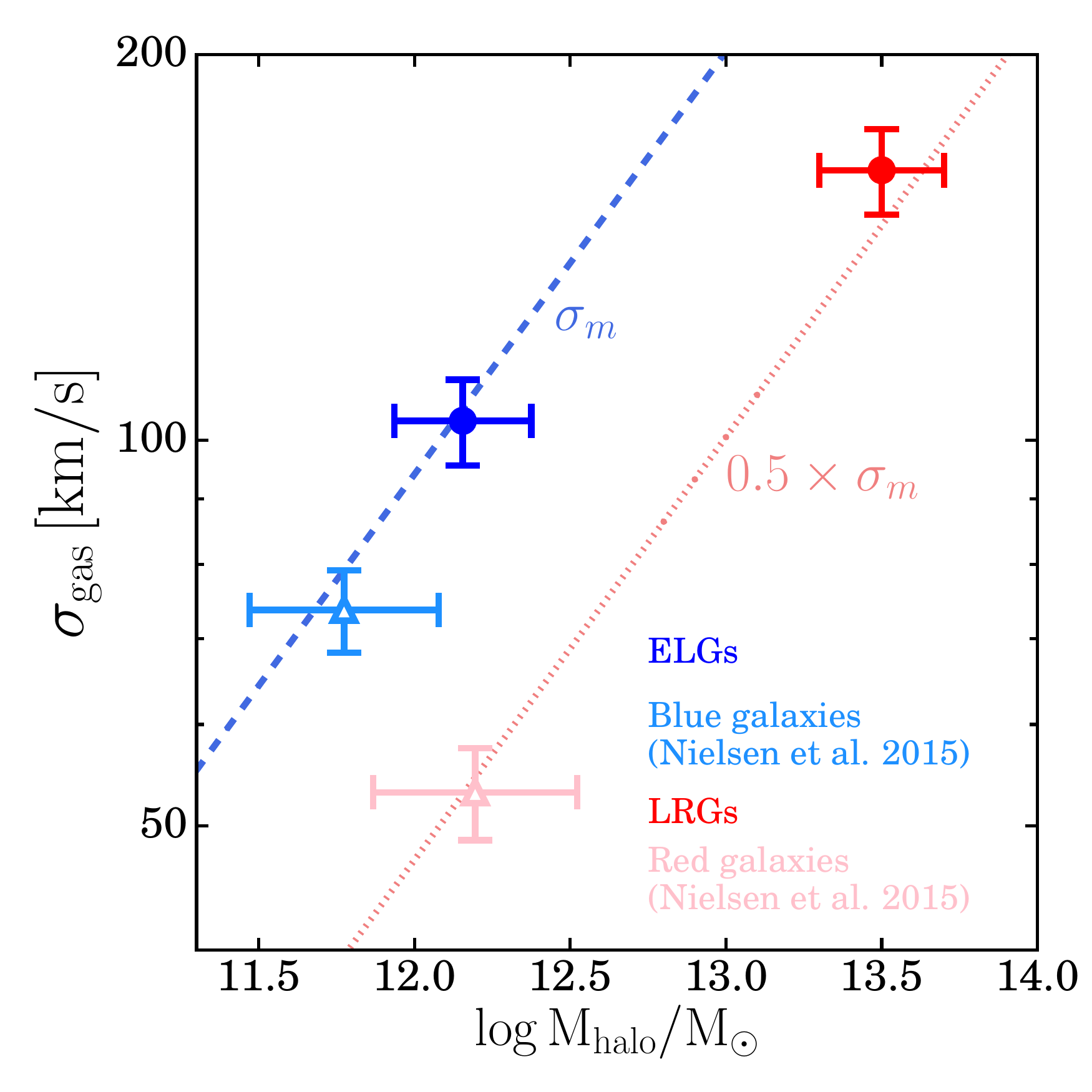}
\caption{
Line-of-sight gas velocity dispersion, $\rm \sigma_{gas}$, as a function of halo
mass and galaxy types. The $\rm \sigma_{gas}$ of ELGs and LRGs are shown by the blue and red data points. We 
find that the gas velocity dispersion around 
ELGs is consistent with the 
velocity dispersion of dark matter particles $\sigma_{m}$ (Equation 2) shown by the blue dashed line, while 
the gas velocity dispersion around LRGs is aligned with $50\%$ of 
the velocity dispersion of dark matter particles shown by the red dotted line. We also show similar measurements around blue and red 
galaxies derived from about 30 individual galaxy-absorber pairs from \citet{Nielsen2015,Nielsen2016}, suggesting that gas in small halos may behave similarly. 
}
\label{}
\end{figure}

Let us first consider the possible mechanisms producing gas
around passive galaxies. 
One possibility is that 
the gas clouds are produced by mass loss from stars in satellite 
galaxies and/or halo stars. In this case, the initial velocities of the clouds 
are expected to follow the virial velocity of the halo, in conflict with the 
observational result. However, in the presence of a diffuse hot halo, the cloud 
velocities are expected to be reduced by the interaction with hot ambient gas. 
Indeed, clouds with higher initial velocities are expected to be destroyed by 
hydrodynamic instabilities, such as Kelvin-Helmholtz and/or Rayleigh-Taylor
instabilities, in shorter timescales, as the instability timescale 
is typically inversely proportional to the cloud velocity, 
$\propto V_{cloud}^{-1}$ \citep[e.g. \S8.5 in][]{Mo2010}.
In addition, gas clouds are also expected to  decelerate due to the ram 
pressure of hot gas. Thus, even if the clouds were produced with 
a velocity dispersion similar to that of the dark matter halo, these two 
mechanisms may work to reduce the gas velocity dispersion observed around 
passive galaxies. Alternatively, the gas clouds around passive galaxies 
may be produced by cooling gas from the hot halo. If the hot gaseous halo 
is static,  the initial velocities of the clouds are expected to be small. 
The gravitational field of the dark matter halo will accelerate the 
gas clouds, but the clouds may get destroyed before they reach high 
velocities, leading to the low cloud velocity dispersion observed.

For star-forming galaxies, a large fraction of the clouds may be produced 
by outflows, as such outflows have been observed ubiquitously 
\citep[e.g.,][]{Heckman2000, Weiner2009,Steidel2010,Martin2012,Bordoloi2014,Rubin2014,Zhu2015,Heckman2015, Chisholm2016}. 
Depending on the initial velocities of the outflows, these clouds may 
propagate to large distances from the halo center, inheriting large 
velocities. To demonstrate this possibility, we perform a simple simulation 
by ejecting particles at a radius of 1 kpc from the center of a dark matter 
halo that is assumed to have  a NFW density profile 
\citep{NFW1996} with total mass of $10^{12} M_{\odot}$ and 
a concentration of $10$. The ejection is assumed to  have 
spherical symmetry and a constant rate over a period of 2 Gyr.
We follow the motion of each ejected particle in the gravitational 
potential well of the halo, and calculate the 
velocity dispersion of ejected particles as a function of the impact 
parameter. The dashed faint blue lines in Figure 5 show the results obtained 
by assuming that the initial velocities following a normal 
distribution with a width 200 km/s, and with a mean velocity 
of 500 km/s, 300 km/s and 200 km/s, respectively. As one can see, 
outflow clouds may contribute significantly to the line of sight 
velocity dispersion, if the initial velocities sufficiently
large. Note that the velocity dispersion obtained from our 
simple model should be considered as upper limits, as other mechanisms, 
such as ram pressure and hydrodynamic instabilities, are expected 
to reduce the velocity dispersion to be observed. In a forthcoming 
paper (Lan et al. in preparation), we will develop a more realistic 
model to test the ideas presented here. 
\begin{figure}
\includegraphics[width=0.5\textwidth]{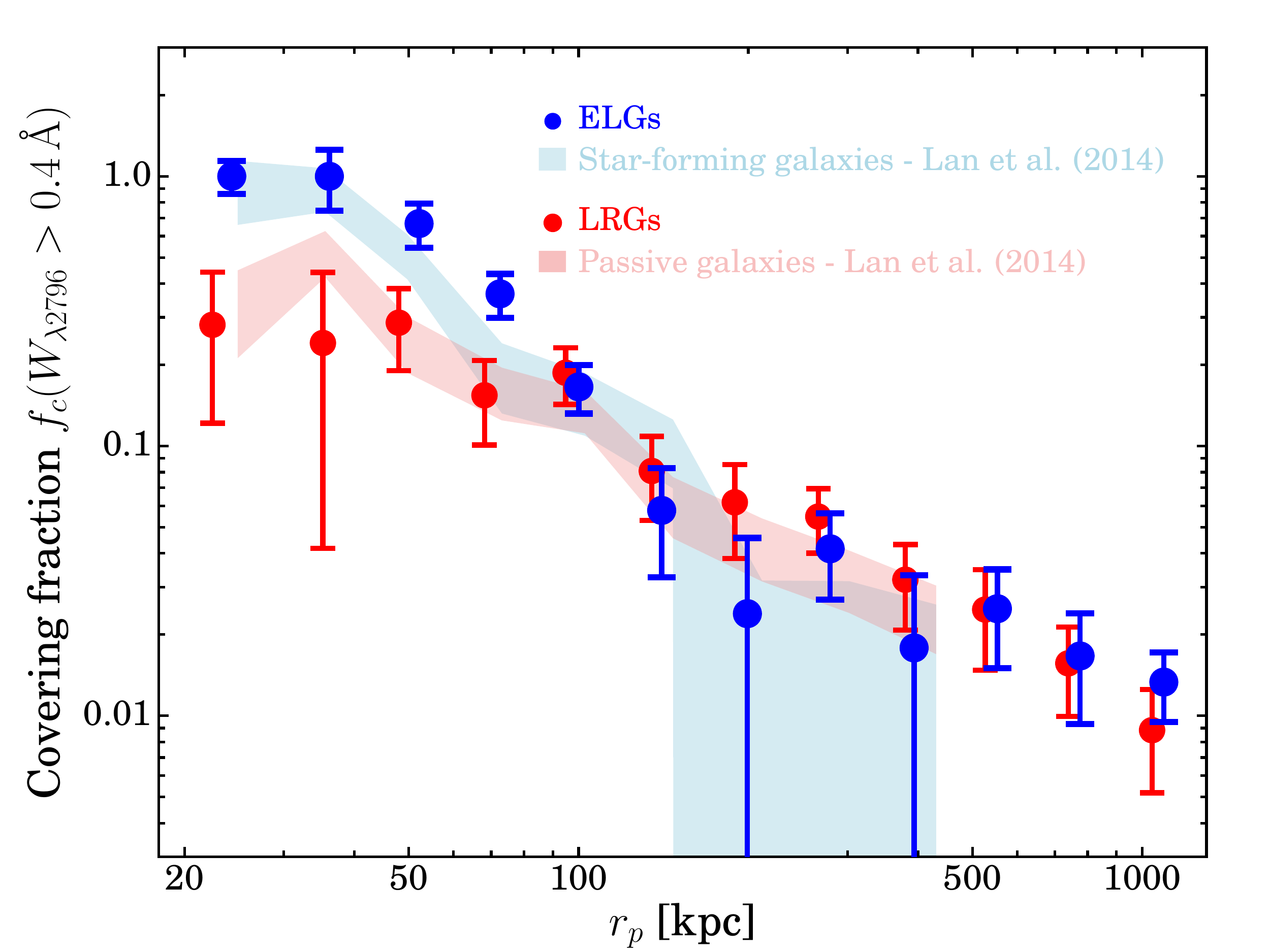}
\caption{
MgII covering fraction around galaxies. The blue data points and red squares show the covering fractions around ELGs and LRGs. The blue and red bands show the covering fractions estimated from \citet{Lan2014} by correlations with individual MgII absorbers and photometric galaxies.
}
\label{fig:cf}
\end{figure}


%
%
\section{Implications}

\subsection{The MgII covering fraction}

The average MgII absorption around galaxies obtained above can be 
converted to the corresponding MgII covering fraction,  $f_{c}$, as 
shown in Equation (1). To do this, we assume that the average MgII absorption 
around galaxies is dominated by strong MgII absorbers ($W_{\lambda2796}>0.4 \rm \, \AA$) 
and that the $W_{\lambda2796}$ distribution around galaxies is similar to the one 
observed towards random quasar sightlines. For simplicity, we approximate the average 
MgII absorption strength as  
\begin{equation}
    \langle W_{\lambda 2796} \rangle(r_{p}) \approx f_{c}(W_{\lambda 2796}>0.4 {\rm \, \AA}, r_{p})\times 
    \hat{W}^{ \rm MgII}_{\lambda 2796},
\end{equation}
where $\langle W_{\lambda 2796} \rangle(r_{p})$ is our average profile,
and $\hat{W}^{ \rm MgII}_{\lambda 2796}$ is the average of
$W_{\lambda 2796}$ obtained from individual absorbers towards 
random quasar sightlines. We use the incidence rate, $d^2N/dWdz$, of individual 
MgII absorbers from \citet{Zhu2013}, and estimate $\hat{W}^{ \rm MgII}_{\lambda 2796}$ using
systems with $W_{\lambda 2796}>0.4 \rm \, \AA$. The corresponding $\hat{W}^{ \rm MgII}_{\lambda 2796}$ is $\sim 1 \, \rm \AA$. 
The estimated covering fraction from Equation (4) is shown in 
Figure~\ref{fig:cf} with the blue and red data points for ELGs and 
LRGs, respectively. Note that the initial covering fraction around ELGs at $<40$ 
kpc exceeds 1, the maximum value for covering fraction, due to that 
MgII absorbers close to star-forming galaxies tend to have higher 
average rest equivalent widths than the global population \citep{Lan2014}. 
For such measurements, the covering fraction is set to be 1. 
The estimated covering fractions are compared with  
that derived from individual absorbers around blue and red 
galaxies (color shaded bands) \citep{Lan2014}. 
These two types of measurements yield consistent results, indicating that 
the assumption and the approximation adopted in Equation (4) are reasonable. 
This demonstrates that the average absorption around ELGs and LRGs is dominated by strong MgII absorbers with $W_{\lambda 2796}>0.4 \rm \, \AA$, with only a negligible contribution from weaker components \citep[see also][]{Prochaska2014}. 
These measurements for LRGs are consistent with \citet{Huang2016}. 
The difference in the covering fractions between ELGs and LRGs 
shows that the MgII gas covering fraction is correlated with star formation activities in addition to halo mass as suggested by \citet{CC2013}. 

We have also performed a simulation to test the viability of using Equation (4) to estimate the covering fraction. To this end, we inject fake absorbers 
with $\hat{W}^{ \rm MgII}_{\lambda 2796}\sim 1 \, \rm \AA$ into the observed individual spectra and run our analysis to recover the input covering fraction. We find that the systematic uncertainty 
is about $20\%$, comparable to the statistical uncertainty.
To better estimate the covering fraction, however, one needs to detect 
individual absorbers in each quasar sightline, instead of using Equation (4). 
We will perform such an analysis in the future.

\begin{figure}
\includegraphics[width=0.5\textwidth]{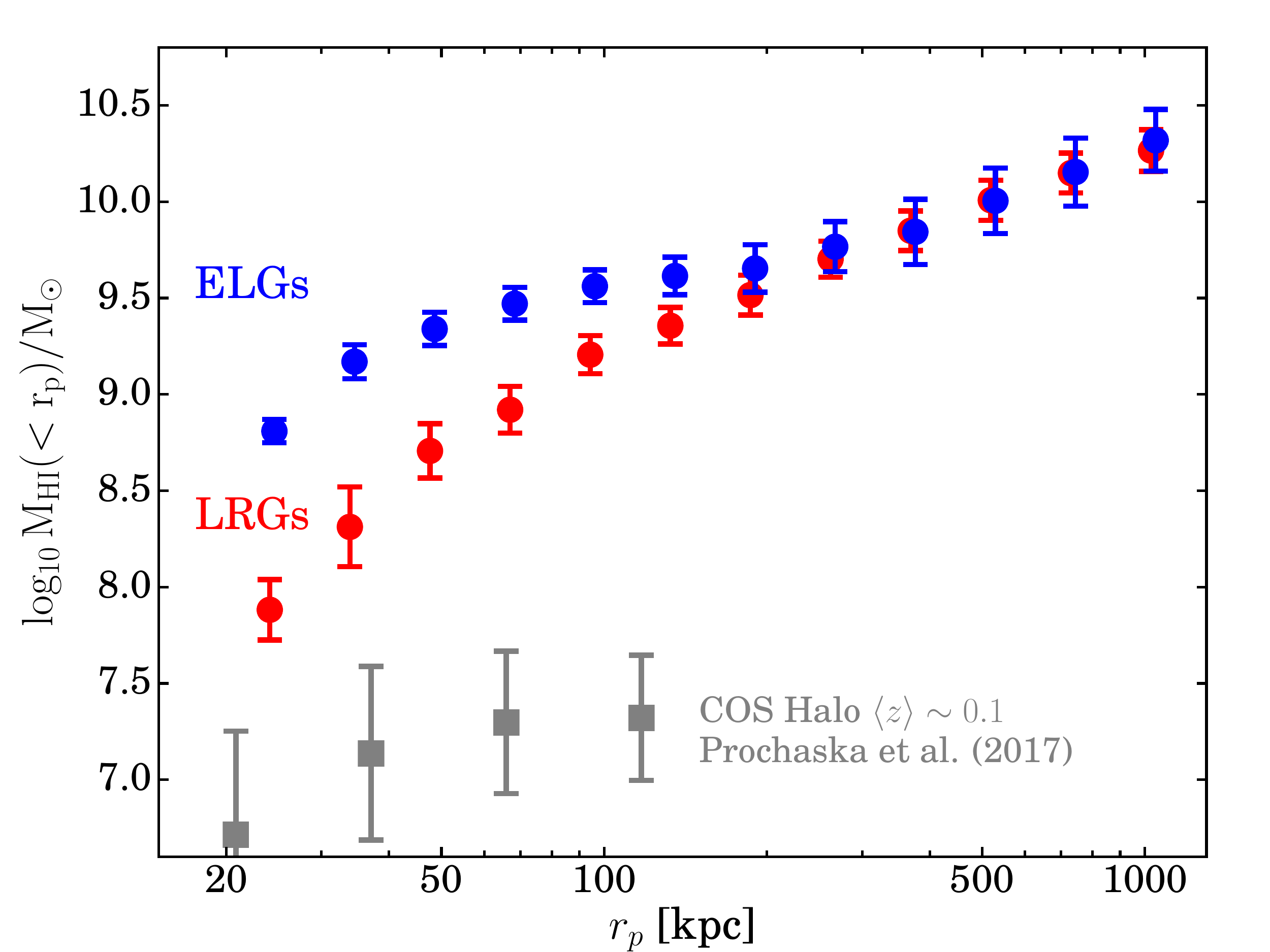}
\caption{
Cumulative neutral hydrogen mass around galaxies as a function of impact parameter. 
}
\label{fig:cumulative_hydrogen}
\end{figure}


\subsection{Mass of circumgalactic neutral hydrogen}

With the covering fraction obtained above, we can estimate the amount of 
hydrogen mass around ELGs and LRGs following \citet{Lan2014},
\begin{equation}
    M_{\rm HI}(r_{p}<r_{p,max})\sim 2\pi \, m_{\rm H} \int_{20 \, \rm kpc}^{r_{p,max}}
    \hat{N}_{\rm HI}\, f_{c}(r_{p})\, r_{p}\, dr_{p}, 
\end{equation}
where $\hat{N}_{\rm HI}$ is the neutral hydrogen column density traced by the 
MgII absorbers. Using the empirical relation between the rest equivalent width of 
MgII and $N_{\rm HI}$ derived by \citet{Lan2017}, we  
obtain $\hat{N}_{\rm HI}\approx 3\times10^{19} \rm cm^{-2}$ for 
$\hat{W}_{\lambda 2796}^{\rm MgII}\sim 1 \rm \, \AA$ 
at redshift 0.8. We perform the integration as a function of impact 
parameters and the enclosed neutral hydrogen mass obtained 
in this way is shown in Figure~\ref{fig:cumulative_hydrogen}. As one can see, 
there is more neutral hydrogen around ELGs than around 
LRGs within $r_p\sim 100\,{\rm kpc}$, but the enclosed mass around the 
two populations becomes comparable at larger scales. 
The enclosed masses within the virial radii of ELGs (200 kpc) 
and LRGs (600 kpc) are $\sim 4\times 10^{9}\, M_{\odot}$ 
and $\sim 9\times 10^{9} \, M_{\odot}$, respectively. 
Compared with the halo masses of ELGs and LRGs, these neutral 
hydrogen masses imply that the HI mass fractions are
   \begin{equation}
    f\sim 
    \left\{\begin{array}{lr}
     10^{-2.5}    & \text{for star-forming galaxies;}\\
     10^{-3.5}  & \text{for passive galaxies.}
     \end{array}
    \right.
    \end{equation}
Thus, the neutral hydrogen mass fraction within the halo around 
ELGs is about 10 times higher than that around LRGs.
As shown in \citet{Lan2017}, the MgII absorption systems used in our analysis are predominantly neutral with $n_{HI}/n_{H}\sim0.9$. Thus, the HI masses given above are approximately the same as the total hydrogen masses traced by MgII absorption lines. 

We can compare the mass of the circumgalactic hydrogen obtained here 
with that at $z\sim 0.1$ obtained from the HST/COS-Halos survey 
\citep[e.g.][]{Stocke2013, Tumlinson2013,Werk2014,Prochaska2017}. 
The COS-Halos results show that the neutral hydrogen is ubiquitously 
around galaxies, with a covering fraction nearly $100\%$ from 10 
kpc to 150 kpc \citep[e.g.,][]{Tumlinson2013,Bordoloi2017}. 
Accounting for the ionization effect,  the estimated total cool CGM mass 
can be as high as $10^{11}\, M_{\sun}$ \citep{Prochaska2017}, more than 10 
times higher than the mass traced by MgII absorption lines. However, 
the hydrogen gas detected in the COS-Halos survey is highly ionized, 
with an ionization correction factor greater than 100. 
To obtain the mass of neutral hydrogen gas detected in the COS-Halos survey, we take the data from \citet[][]{Prochaska2017} and estimate the median 
$N_{\rm HI}$ for all types of galaxies as a function of impact parameters. 
We then calculate the neutral hydrogen mass using Equation (5)
and assuming a covering fraction of $100\%$ from 20 kpc to 150 kpc. 
We find that the total mass of neutral hydrogen detected by COS/Halos is only 
about $10^{7.5} \, M_{\sun}$, a factor of about 100 lower than the amount 
around ELGs shown in Figure~\ref{fig:cumulative_hydrogen}. 
This result indicates that the bulk of neutral hydrogen around galaxies is 
traced by strong MgII absorbers. The difference of the circumgalactic 
neutral hydrogen between ELGs and local galaxies may be caused by 
two factors. First, the number of COS-Halos sightlines is 
not big enough to properly sample the cool dense clouds with 
relatively high HI column density and low covering 
fraction. Second, the amount of neutral 
hydrogen around galaxies decreases significantly from $z\sim 1$ 
to 0.1. At the moment, it is unclear which of the two is the right 
reason.

Our mass estimates also allow us to quantify the cosmic mass density of 
neutral hydrogen contributed from the CGM of ELGs ($r_{p}<200$ kpc). 
To do this, we assume that all the star-forming galaxies with 
$M_{*}>10^{10} M_{\odot}$ have similar gas profiles as ELGs 
and use the number density of star-forming galaxies, 
$\rm n (SFR>2 \, M_{\odot}/yr)\sim 10^{-2.6} \, Mpc^{-3}$ at $z\sim0.8$ from 
\citet{Moustakas2013}. 
The cosmic mass density of neutral 
hydrogen around ELGs at $z\sim0.8$ estimated in this way is 
\begin{equation}
    \Omega_{\rm HI}^{\rm ELGs}=\frac{n\times M_{\rm HI}}{\rho_{\rm crit}} \approx 1\times10^{-4}.
\end{equation}
Using the incidence rate of individual MgII absorbers and the empirical 
relation between $N_{\rm HI}$ and $W_{\lambda 2796}$, 
previous studies \citep{Kacprzak2011, Menard2012, Lan2017} estimated the total mass density 
of neutral hydrogen traced by MgII absorbers 
to be $\Omega_{\rm HI}^{\rm MgII} \approx 1.5\times10^{-4}$ around 
redshift 1. This suggests that more than $60\%$ of the individual 
MgII absorbers observed towards random quasar sight-lines are 
actually associated with the CGM of massive star-forming galaxies. 
This result is also consistent with the fact that on average, 
MgII absorbers have approximately solar metallicity at 
$z\sim 1$ \citep{Lan2017}.

\subsection{Outflow rate}

As shown in Figure 8, the amount of neutral hydrogen mass within 
100 kpc around ELGs exceeds from that around LRGs by about 
$2\times10^{9} \rm \, M_{\odot}$.
Assuming that this mass difference is due to mass ejected from star-forming galaxies, we can estimate an outflow rate. We assume that the outflow material
has a typical velocity of 200 km/s and typically travels a distance of 
$\sim 100$ kpc. The corresponding time scale is then about 500 Myr. 
If the excess mass around ELGs is assumed to be ejected about 500 Myr ago, 
we can estimate the outflow rate to be
\begin{equation}
    \rm \dot{M}_{out}=dM/dt \sim 2\times10^{9}\, M_{\odot}/500 \, Myr \sim 4 \rm \, M_{\odot}\,  yr^{-1}. 
\end{equation}
This estimated value should be considered as a lower limit, as about 80\% of the 
excess mass is within 60 kpc, where the outflow velocity is lower than the 
typical outflow velocity. In addition, only cool gas clouds traced by MgII 
absorption are included, and a fraction of outflow materials may have already
fallen back to the galaxies or been destroyed. Assuming that the SFR of the galaxies does not 
evolve significantly over 500 Myr, we can obtain a constraint on the minimum 
outflow loading factor:
\begin{equation}
    \frac{\rm \dot{M}_{out}}{\rm SFR}>4/8 \sim 0.5,
\end{equation}
a value similar to the one obtained from the blue-shifted absorption lines 
towards galaxies \citep[e.g.][]{Rubin2014}.

\subsection{Comparison with previous studies}
The absorption profile for ELGs at redshift 0.8 can be 
compared with that around Lyman break galaxies (LBGs) at redshift 2.2. 
To do this, we use the CII$\lambda 1334$ absorption strength 
obtained from \citet{Steidel2010} and convert it to the corresponding 
MgII absorption using  $\rm W_{MgII,all}\sim 3\times \, W_{CII\lambda1334}$ 
as given in \citet{Lan2017}. We find that at
$r_p\sim 50$ kpc, the absorption around LBGs ($\sim 2 \rm\, \AA$) appears 
to be stronger than that around ELGs ($\sim 1 \rm\, \AA$), indicating a 
possible evolution of gas around star-forming galaxies. 
Unfortunately, the LBG data is still too 
uncertain to provide quantitative constraints on the evolution. 
We also compare our measurements with similar measurements from  \citet{Rubin2018} and 
\citet{Bordoloi2011}, both using spectra of galaxies as background 
sources. These measurements appear to be lower than ours but have 
large uncertainties. The difference could be due to that (1) their 
foreground star-forming galaxies have, on average, lower stellar masses 
and (2) they use extended objects as background sources, which 
may probe a larger region per line of sight as compared with quasars 
\citep[see,][for a discussion]{Bordoloi2011}.

\subsection{Origins of the MgII/FeII absorption gas}
\label{ssec_origins}

Our results show that the properties of the cool gas around ELGs and LRGs are different. 
The gas profile between 10 kpc to 1 Mpc
around LRGs is consistent with the NFW profile for dark matter distribution, as shown in \citet{Zhu2014}, which can be described roughly by a power law.
In contrast, the gas profile around ELGs is 
much steeper at impact parameters below 100 kpc, although it is 
comparable to that of LRGs at larger impact parameters.
The properties of the absorbing gas is also found to be correlated with 
the star-formation activity and the azimuthal angle, 
with the absorption strength being stronger around ELGs with 
higher SFR and for sight-lines closer to the minor axes of the galaxies. 
In addition, the gas around star-forming galaxies appear to be stirred up 
in their halos than that around passive galaxies after normalizing 
the effect of halo mass. These results clearly have important 
implications for the origins of the absorption gas. In what follows, 
we discuss two possible mechanisms.

\textbf{Gas associated with outflows} - We argue that most of the cool gas around 
star-forming galaxies traced by MgII/FeII absorption lines within the halos is 
associated with outflows. Star-forming galaxies are known to eject gas via 
outflows. Evidence for such flows can be seen directly from the blue-shifted 
absorption lines in the galaxy spectra, i.e. through the so-called 
down-the-barrel observations \citep[e.g.,][]{Weiner2009,Steidel2010,Martin2012,
Bordoloi2014,Rubin2014,Zhu2015,Chisholm2016}. 
However, the lack of spatial information in such observations makes it 
difficult to infer how far the gas associated with outflows can propagate 
into the halos. 
The azimuthal angle dependence of gas absorption found here is consistent with the scenario that the CGM is enriched by outflows. 
The azimuthal angle dependence is difficult to be explained by gas inflows and/or gas associated with 
satellite galaxies, because the gas distribution produced by 
such mechanisms is not expected to be aligned with the minor
axes of galaxies. Indeed, although the accretion of material from the cosmic web 
into dark matter halos is anisotropic \citep[e.g.,][]{Shi2015}, the alignment 
between the principal axes of late type galaxies and the cosmic web is 
found to be weak \citep[e.g.,][]{Zhang2013}, so is the alignment 
of late type galaxies with the distribution of satellite galaxies 
\citep[e.g.,][]{Yang2006}.
We conclude, therefore, that the CGM around 
star-forming galaxies are produced and enriched by outflows driven 
by star formation in the galaxies.

The outflow gas traced by MgII around massive star-forming galaxies
seems to have a characteristic scale of about 50-100 kpc
\citep[see also][]{Bordoloi2011}. This indicates that the bulk 
of the outflow materials in the cool phase cannot travel much 
farther than $\sim100$ kpc from the central galaxies. 
This characteristic scale is an important observational constraint 
on any models of galactic outflows.

\textbf{Gas associated with halos} - We argue that, in addition to 
galactic outflows, a fraction of the MgII/FeII absorbing gas 
originates from mechanisms associated with halos of galaxies. 
This is motivated by the fact that a non-negligible amount of 
cool gas is found around LRGs, a population with no significant
star formation activity over the past 1-2 Gyr or even longer \citep[e.g.,][]{Barber2007, Gauthier2011}. 
Furthermore, the cool gas around LRGs has a relatively low velocity 
dispersion, and a distribution consistent with the NFW matter 
distribution without significant azimuthal dependence, all 
consistent with a halo origin. In the following, we discuss two 
possible mechanisms that may produce absorbing clouds in halos 
of galaxies. These mechanisms may also operate around star-forming 
galaxies, given that all galaxies are surrounded by extended halos. 

One possibility is that the absorbing clouds originate from satellite 
galaxies. These galaxies are known to roughly trace the dark matter 
distribution \citep[e.g.,][]{Lin2004}, which may explain the 
observed gas density profile around LRGs. Before being accreted
into their host halos, star-forming satellite galaxies have cool 
gas halos enriched by outflows, as discussed above. After merging into 
its host halo, a satellite may lose a fraction of the cool CGM  
due to ram pressure and/or tidal stripping, but it may still retain
part of the CGM. Before being destroyed by hydro-dynamical instabilities 
and/or heat evaporation, both the remaining and stripped circumgalactic 
gas can contribute to the cool gas seen around LRGs. 
In addition, gas clouds may also be produced by the mass loss from evolved 
stars in the satellites. Under the assumption that satellite galaxies 
can retain {\it all of their cool CGM} without being destroyed or removed 
by any mechanisms, it seems possible to explain the observed MgII 
gas profiles observed around galaxy groups and LRGs by the gas 
associated with satellites \citep[e.g.,][]{Bordoloi2011,Huang2016}. 
However, if most of the absorbing gas were still bound to satellite 
galaxies, the gas kinematics would follow that of the satellites, 
and therefore that of dark matter \citep[e.g.,][]{More2011}, 
inconsistent with the observed gas kinematics. 
On the other hand, lower cloud velocity dispersion is expected 
for the clouds unbound to satellites, as ram pressure can 
decelerate them and hydrodynamic instabilities tend to destroy 
faster moving clouds in shorter time scales (see \S\ref{ssec_kinematics}).  

The cooling gas condensed from hot diffuse halos through thermal 
instability is another possible source for the MgII gas around LRGs. 
This scenario was first proposed by \citet{Mo1996}, with an extension 
developed by \citet{Maller2004}, and it assumes that a significant 
amount of hot gas within the cooling radius will 
condense into cool gas clouds. Given that the cooling time strongly 
depends on gas metallicity, it is expected that the high metallicity 
regions in the hot halos will cool first and produce gas clouds with 
high metal content. In addition, the condensed cool gas clouds are 
expected to have velocities modulated by the gravity of the systems 
and by the ram pressure of the hot halos, probably leading to low 
velocity dispersion. These two properties, high metallicity and 
low velocity dispersion, are consistent with the observed properties 
of the MgII gas, as shown in this paper and in \citet{Lan2017}. 
However, it is still unclear if this mechanism can produce the 
observed gas distribution. 

It is possible that all the mechanisms discussed above can contribute 
to the cool gas around LRGs. The question that remains unsolved 
is their relative importance. By studying the contribution of each 
mechanism in detail through analytic and numerical models, 
it is possible to disentangle the origins of the cool gas around LRGs without
the complications introduced by outflows. Thus, the gas properties 
around passive galaxies may provide a cleaner test bed for CGM models 
based on the halo origin than that around star-forming galaxies. 

%
%
\section{Summary}

The circumgalactic medium is expected to contain important information about 
gas accretion into galaxies, as well as gas flow processes that drive the evolution of 
galaxies. To reveal the imprints of these processes, we measure the distribution and
kinematics of cool gas around star-forming and passive galaxies, using MgII/FeII absorption 
properties extracted from the flux decrements in the spectra of background quasars
around $\sim200,000$ ELGs and $\sim800,000$ LRGs. Our findings can be summarized as follows:

\begin{enumerate}

    \item ELGs appear to be surrounded by more cool gas than LRGs within 100 kpc despite being much less massive.
    The MgII and FeII absorption around ELGs is about 5-10 times stronger than around LRGs, and  
    is stronger around ELGs with higher SFR. 
    At larger scales, in contrast, the absorption around ELGs 
    and LRGs has a similar strength and decreases with impact parameters following a power 
    law, $r_{p}^{-1}$.
    
    \item The metal absorption distribution around ELGs is anisotropic; 
    for impact parameters below $\sim 100$ kpc, the metal absorption along the 
    minor axes of ELGs is, on average, about two times stronger than 
    that along the major axis. This indicates that a significant fraction of the 
    absorbing gas, $\sim 2/3$, is probably generated by outflows from the galaxies.
    
    \item The line-of-sight gas velocity dispersion $\sigma_{\rm gas}$ around ELGs and LRGs 
    within 200 kpc is measured to be about 100 km/s and 160 km/s from the absorption line 
    widths, respectively. Comparing these with the expected velocity dispersion of dark 
    matter particles, $\sigma_{\rm m}$, in the host halos, we find that the gas moves 
    differently around ELGs and LRGs, with the gas velocity bias  
    \begin{equation}
    {\sigma_{gas}\over \sigma_{m}}\sim 
    \left\{\begin{array}{lr}
     1    & \text{for ELGs;}\\
     0.5  & \text{for LRGs.}
     \end{array}
    \right.
    \end{equation}

    \item We infer the covering fraction of individual MgII absorbers  and 
    estimate the amount of neutral hydrogen around ELGs and LRGs based on an 
    empirical relation between MgII strength and neutral hydrogen column 
    density $\rm N_{HI}$. Within 100 kpc, the neutral hydrogen mass around ELGs 
    is about $2\times10^{9} \, M_{\odot}$, about a factor of 3 more than 
    that around LRGs. Assuming this difference is contributed by outflow materials, 
    we constrain the minimum outflow loading factor to be about 0.5.
\end{enumerate}

Our results demonstrate that the properties of cool gas around ELGs and LRGs within 
100 kpc are significantly different in terms of their metal absorption profiles, 
gas abundance, and kinematics. This correlation between the gas properties and 
the SFR of galaxies is consistent with the picture that the CGM is enriched by 
metal-rich gas ejected via galactic outflows associated with recent star 
formation activities. Thus, the observed properties of the cool gas around 
star-forming galaxies can provide essential constraints on the feedback 
processes in galaxy formation. In a forthcoming paper (Lan et al. in preparation), 
we will use these results, together with a semi-analytic model for gaseous
galactic halos, to constrain how the gas flow from galaxies can affect 
the circumgalactic media.  

Our results also demonstrate the potential to measure the gas properties around 
galaxies with unprecedented precision by combining large samples and powerful
statistical technique. Indeed, the large sample provided by SDSS already makes 
it possible to measure both the gas absorption towards 
ELGs \citep['down-the-barrel' observation,][]{Zhu2015} and the gas absorption 
around ELGs robustly. With the advents of even larger and deeper spectroscopic 
samples of galaxies, such as DESI \citep{Schlegel2011,Levi2013}, 
PFS \citep{Takada2014}, Euclid \citep{Amiaux2012}, SDSS-V \citep{Kollmeier}, 
4MOST \citep{4most} and MOONs \citep{Moons}, we will be able to use more absorption 
species (MgII, FeII, CIV, etc) to probe the cosmic evolution of the circumgalatic gas-galaxy 
interaction, eventually obtaining a complete picture of baryon cycle in galaxy 
formation and evolution.  In addition, it is also possible to use the 
galaxy-metal line cross-correlation to study the large-scale structure 
and even to detect the BAO signal \citep[e.g.,][]{Blomqvist2018}. 

\acknowledgements

We thank Brice M\'enard for his substantial suggestions on the manuscript, J.Xavier Prochaska for useful discussions, Nikole Nielsen for providing the kinematics data of her MgII absorber catalog, Dustin Lang for his helps on calculating the azimuthal angle, and Guangtun Zhu for sharing the NMF eigen-spectra of SDSS quasars. We also want to thank the anonymous referee for the constructive report.
HJM acknowledges support from NSF AST-1517528,  and from 
National Science Foundation of China (grant Nos. 11673015, 11733004). Kavli IPMU is supported by World Premier International
Research Center Initiative of the Ministry of Education, Japan.

Funding for the Sloan Digital Sky Survey IV has been provided by the Alfred P. Sloan Foundation, the U.S. Department of Energy Office of Science, and the Participating Institutions. SDSS acknowledges support and resources from the Center for High-Performance Computing at the University of Utah. The SDSS web site is www.sdss.org.

SDSS is managed by the Astrophysical Research Consortium for the Participating Institutions of the SDSS Collaboration including the Brazilian Participation Group, the Carnegie Institution for Science, Carnegie Mellon University, the Chilean Participation Group, the French Participation Group, Harvard-Smithsonian Center for Astrophysics, Instituto de Astrofísica de Canarias, The Johns Hopkins University, Kavli Institute for the Physics and Mathematics of the Universe (IPMU) / University of Tokyo, Lawrence Berkeley National Laboratory, Leibniz Institut für Astrophysik Potsdam (AIP), Max-Planck-Institut für Astronomie (MPIA Heidelberg), Max-Planck-Institut für Astrophysik (MPA Garching), Max-Planck-Institut für Extraterrestrische Physik (MPE), National Astronomical Observatories of China, New Mexico State University, New York University, University of Notre Dame, Observatório Nacional / MCTI, The Ohio State University, Pennsylvania State University, Shanghai Astronomical Observatory, United Kingdom Participation Group, Universidad Nacional Autónoma de México, University of Arizona, University of Colorado Boulder, University of Oxford, University of Portsmouth, University of Utah, University of Virginia, University of Washington, University of Wisconsin, Vanderbilt University, and Yale University.

The Legacy Surveys consist of three individual and complementary projects: the Dark Energy Camera Legacy Survey (DECaLS; NOAO Proposal ID \# 2014B-0404; PIs: David Schlegel and Arjun Dey), the Beijing-Arizona Sky Survey (BASS; NOAO Proposal ID \# 2015A-0801; PIs: Zhou Xu and Xiaohui Fan), and the Mayall z-band Legacy Survey (MzLS; NOAO Proposal ID \# 2016A-0453; PI: Arjun Dey). DECaLS, BASS and MzLS together include data obtained, respectively, at the Blanco telescope, Cerro Tololo Inter-American Observatory, National Optical Astronomy Observatory (NOAO); the Bok telescope, Steward Observatory, University of Arizona; and the Mayall telescope, Kitt Peak National Observatory, NOAO. The Legacy Surveys project is honored to be permitted to conduct astronomical research on Iolkam Du'ag (Kitt Peak), a mountain with particular significance to the Tohono O'odham Nation.

NOAO is operated by the Association of Universities for Research in Astronomy (AURA) under a cooperative agreement with the National Science Foundation.

This project used data obtained with the Dark Energy Camera (DECam), which was constructed by the Dark Energy Survey (DES) collaboration. Funding for the DES Projects has been provided by the U.S. Department of Energy, the U.S. National Science Foundation, the Ministry of Science and Education of Spain, the Science and Technology Facilities Council of the United Kingdom, the Higher Education Funding Council for England, the National Center for Supercomputing Applications at the University of Illinois at Urbana-Champaign, the Kavli Institute of Cosmological Physics at the University of Chicago, Center for Cosmology and Astro-Particle Physics at the Ohio State University, the Mitchell Institute for Fundamental Physics and Astronomy at Texas A\&M University, Financiadora de Estudos e Projetos, Fundacao Carlos Chagas Filho de Amparo, Financiadora de Estudos e Projetos, Fundacao Carlos Chagas Filho de Amparo a Pesquisa do Estado do Rio de Janeiro, Conselho Nacional de Desenvolvimento Cientifico e Tecnologico and the Ministerio da Ciencia, Tecnologia e Inovacao, the Deutsche Forschungsgemeinschaft and the Collaborating Institutions in the Dark Energy Survey. The Collaborating Institutions are Argonne National Laboratory, the University of California at Santa Cruz, the University of Cambridge, Centro de Investigaciones Energeticas, Medioambientales y Tecnologicas-Madrid, the University of Chicago, University College London, the DES-Brazil Consortium, the University of Edinburgh, the Eidgenossische Technische Hochschule (ETH) Zurich, Fermi National Accelerator Laboratory, the University of Illinois at Urbana-Champaign, the Institut de Ciencies de l'Espai (IEEC/CSIC), the Institut de Fisica d'Altes Energies, Lawrence Berkeley National Laboratory, the Ludwig-Maximilians Universitat Munchen and the associated Excellence Cluster Universe, the University of Michigan, the National Optical Astronomy Observatory, the University of Nottingham, the Ohio State University, the University of Pennsylvania, the University of Portsmouth, SLAC National Accelerator Laboratory, Stanford University, the University of Sussex, and Texas A\&M University.

BASS is a key project of the Telescope Access Program (TAP), which has been funded by the National Astronomical Observatories of China, the Chinese Academy of Sciences (the Strategic Priority Research Program "The Emergence of Cosmological Structures" Grant \# XDB09000000), and the Special Fund for Astronomy from the Ministry of Finance. The BASS is also supported by the External Cooperation Program of Chinese Academy of Sciences (Grant \# 114A11KYSB20160057), and Chinese National Natural Science Foundation (Grant \# 11433005).

The Legacy Survey team makes use of data products from the Near-Earth Object Wide-field Infrared Survey Explorer (NEOWISE), which is a project of the Jet Propulsion Laboratory/California Institute of Technology. NEOWISE is funded by the National Aeronautics and Space Administration.

The Legacy Surveys imaging of the DESI footprint is supported by the Director, Office of Science, Office of High Energy Physics of the U.S. Department of Energy under Contract No. DE-AC02-05CH1123, by the National Energy Research Scientific Computing Center, a DOE Office of Science User Facility under the same contract; and by the U.S. National Science Foundation, Division of Astronomical Sciences under Contract No. AST-0950945 to NOAO.



\end{document}